\documentclass[twocolumn,multicolumn,aps,prb,showpacs]{revtex4}
\usepackage{graphicx}
\usepackage{dcolumn}
\usepackage{bm}
\usepackage{color}
\usepackage{latexsym}
\usepackage[T1]{fontenc}
\usepackage{amsmath}

\input{epsf}

\begin{document}

\preprint{APS/123-QED}

\date{\today}

\title{Vortex creep and critical current densities in superconducting (Ba,K)Fe$_{2}$As$_{2}$ single crystals}

\author{M. Ko\'{n}czykowski}
\affiliation{Laboratoire des Solides Irradi\'{e}s, CNRS-UMR 7642 \& CEA-DSM-IRAMIS, Ecole Polytechnique, F 91128 Palaiseau cedex, France}
\author{C.~J.~van der Beek}
\affiliation{Laboratoire des Solides Irradi\'{e}s, CNRS-UMR 7642 \& CEA-DSM-IRAMIS, Ecole Polytechnique, F 91128 Palaiseau cedex, France}
\author{M. A. Tanatar}
\affiliation{The Ames Laboratory, Ames, IA 50011, U.S.A.}
\author{Huiqian Luo and Zhaosheng Wang}
\affiliation{National Laboratory for Superconductivity, Institute of Physics and National Laboratory for Condensed Matter Physics, Chinese Academy of Sciences,
Beijing 100190, China}
\author{Bing Shen, Hai Hu Wen}
\affiliation{National Laboratory of Solid State Microstructures and Department of Physics, Nanjing University, Nanjing 210093, China}
\author{R.~Prozorov}
\affiliation{The Ames Laboratory, Ames, IA 50011}
\affiliation{Department of Physics \& Astronomy, Iowa State University, Ames, IA 50011, U.S.A.}

\date{\today}

\begin{abstract}
The surprisingly rapid relaxation of the sustainable current density in the critical state of single crystalline Ba$_{1-x}$K$_{x}$Fe$_{2}$As$_{2}$ is investigated for magnetic fields oriented parallel to the $c$-axis and to the $ab$--plane respectively. Due to the inadequacy of standard analysis procedures developed for flux creep in the high temperature superconducting cuprates, we develop a simple, straightforward data treatment technique that reveals the creep mechanism and the creep exponent $\mu$.  At low magnetic fields, below the second magnetization peak, $\mu$ varies only slightly as function of temperature and magnetic flux density $B$.  From the data, we determine  the temperature- and field dependence of the effective activation barrier for creep. At low temperatures, the measured current density approaches the zero--temperature critical current density (in the absence of creep) to within a factor 2, thus lending credence to earlier conclusions drawn with respect to the pinning mechanism. The comparable values of the experimental screening current density and the zero-temperature critical current density reveals the limited usefulness of the widely used ``interpolation formula''. 
\end{abstract}

\maketitle

\section{Introduction}

Recently, the measurement of the critical current density $j_{c}$ of superconducting iron-based compounds has been recognized as a useful tool for the characterization of microscopic\cite{Kees,Kees1} and nanoscale\cite{Kees,Sultan} disorder in these materials. However, the study of vortex pinning and the critical current density is compromised by surprisingly large thermally activated flux creep\cite{Prozorov2008,HuangYang2008,BingShen2010} in the Bean critical state.\cite{Bean62,Zeldov94} In some materials, such as single crystalline (Ba,K)Fe$_{2}$As$_{2}$, the logarithmic creep rate $S \equiv - d\ln j / d \ln t$ of the sustainable current density $j$ approaches that previously measured in some of the high temperature cuprate superconductors.\cite{Yeshurun88,Malozemoff91,Konczykowski} Sizeable creep rates influence the magnitude, and, potentially,  the temperature and flux density-dependence $j(T,B)$, thereby compromising the analysis of fundamental vortex pinning mechanisms in the iron-based superconductors\cite{Kees,Kees1} and the understanding of possible phase transitions of the vortex ensemble.\cite{Kees1,SalemSugui,secondpeak} 

A detailed analysis of flux creep in the iron-based superconductors is therefore justified.  In the cuprates, such analysis has unveiled the non-logarithmic nature of vortex creep, a direct consequence of the non-linearity of the relevant potential barrier $U(j)$ opposing thermally activation as function of the driving force $Bj$. In turn, this nonlinear behavior arises from the elasticity of the vortex ensemble, {\em i.e.}, the fact that this can be deformed continuously on very different length scales. In contrast to single-particle creep,\cite{Anderson62} the relevant activation barrier does not depend algebraically on the driving force, but, rather, increases steeply at low driving forces because of the nonlinear increase of the size of the critical nucleus.\cite{Feigelman89,Fisher89,Nelson92}  In general, the size of the critical nucleus increases as an inverse power-law in $j$, leading to the well-known relation\cite{Feigelman89,Fisher89,Nelson92} 
\begin{equation}
U(j) = U_{c}\left(\frac{j_{c}}{j}\right)^{\mu}.
\label{eq:cc}
\end{equation}
The value of the creep exponent $\mu$ depends on the dimensionality of the critical nucleus as well as that of the elastic manifold as a whole.\cite{Feigelman89} In the case of single crystalline iron-based superconductors to be considered here, the latter can be either a single one-dimensional (1D) vortex line, or the 3D vortex ensemble. The law (\ref{eq:cc}) does not extrapolate to a zero activation barrier at large driving force; one therefore frequently resorts to the so-called interpolation formula,\cite{Fisher89,Nelson92}
\begin{equation}
U(j) = U_{c} \left[ \left(\frac{j_{c}}{j}\right)^{\mu} -1 \right].
\label{eq:interpolation}
\end{equation}

Among the various methods to experimentally establish the $U(j)$--dependence,\cite{Koch89,Malozemoff91,Maley90,Schnack93,Konczykowski93,Konczykowski95} the analysis of the logarithmic time dependence of $1/S$ is the most reliable:\cite{Konczykowski93,Konczykowski95} for Eq.~(\ref{eq:interpolation}), one has $1/S \propto \mu \ln[(t_{0} + t) / \tau]$, with $t_{0}$ a constant determined by transients at the onset of relaxation, and $\tau = (\Lambda j_{c}/E_{0})\left(k_{B}T/U_{c}\right)$ a normalization time determined by the barrier magnitude, a factor $E_{0}$ (with dimension of electric field) related to the details of the creep mechanism, and a sample ``inductance'' $\Lambda = \mu_{0}a^{2}$ ($a$ is a relevant sample dimension).\cite{vdBeek92II,Schnack92} 
An alternative ``dynamic'' method based on the dependence of the sustainable current density on the sweep-rate $\dot{H}_{a}$ of the applied magnetic field in magnetic hysteresis loop measurements was used in Refs.~\onlinecite{HuangYang2008,BingShen2010,Schnack93}, and \onlinecite{Griessen94,HHWen95,HHWen97}.  In principle, the product $(T/Q) d \ln j/dT$, with $Q \equiv d \ln j/d \ln \dot{H}_{a}$, directly yields $\mu$.\cite{HuangYang2008,BingShen2010,Griessen94,HHWen95,HHWen97,HHWen2001} 
The potential of local magnetic measurements of the magnetic induction $B$ was exploited by Abulafia {\em et al.}, \cite{Abulafia95} who reconstructed the current-voltage characteristics of YBa$_{2}$Cu$_{3}$O$_{7-\delta}$ single crystals from Amp\`{e}re's law $j \sim \mu_{0}^{-1}dB/dx$ for the current density  and Faraday's law 
\begin{equation}
E(x) = - \int_{0}^{x} (dB/dt) dx^{\prime}
\label{eq:E}
\end{equation}
for the electric field (integrating from the sample centre to its perimeter). 

As opposed to the cuprates, the dynamic range over which creep data can be collected in the iron--based superconductors is insufficient to reliably determine $S(\ln t)$, $Q(\ln \dot{H}_{a})$, or  the curvature of $E(j) = E_{0} \exp ( -U(j) / k_{B}T )$. Moreover, the factor $d \ln j/dT \propto [1- ( k_{B}T/U_{c} )(dU_{c}/dT)]$ used in the dynamic method admixes the temperature dependence $U_{c}(T)$ with the creep exponent $\mu$.  As a result, it is difficult to distinguish between a logarithmic $U \sim U_{c}\ln(j_{c}/j)$ with a temperature-dependent $U_{c}$, and a so-called ``negative $\mu$''-type barrier, $U = U_{c}[ 1 - (j/j_{c})^{|\mu|}]$.\cite{HHWen97,HHWen2001} One should therefore resort to other constructions, such as that proposed by Maley {\em et al.},\cite{Maley90,vdBeek92II,HuangYang2008,SalemSugui} which combine the results of relaxation measurements at different $T$. 

Below we show that even though this method is fraught with shortcomings, it can be suitably adapted to yield reliable results. In particular, a plot of the average $\langle -  k_{B}T d\ln(|dj/dt|)/dj \rangle$ vs. the average $\langle j \rangle$ unambiguously yields the curvature of the $U(j)$--relation and therefore the flux creep mechanism. Applying this to single crystalline (Ba,K)Fe$_{2}$As$_{2}$, one finds, first of all, that the creep rate is insufficient to qualitatively modify the field dependence $j(B)$ as obtained from magnetic hysteresis. Second, the low--temperature sustainable current density $j$ is, typically, less than a factor of 2 lower than the value the critical current density $j_{c}$ would have in the absence of creep. Finally, the obtained $\mu$-values indicate that the same creep mechanism is relevant for all temperatures up to the transformation of the vortex ensemble at the so-called ``second magnetization peak''.\cite{secondpeak} We discuss the $\mu$--value and its increase as function of magnetic field in terms of the interplay of strong pinning by nm-scale heterogeneities and weak collective pinning by atomic scale point defects in the material.

\begin{figure}[t]%
\vspace{-5mm}
\includegraphics[width=1.1\linewidth]{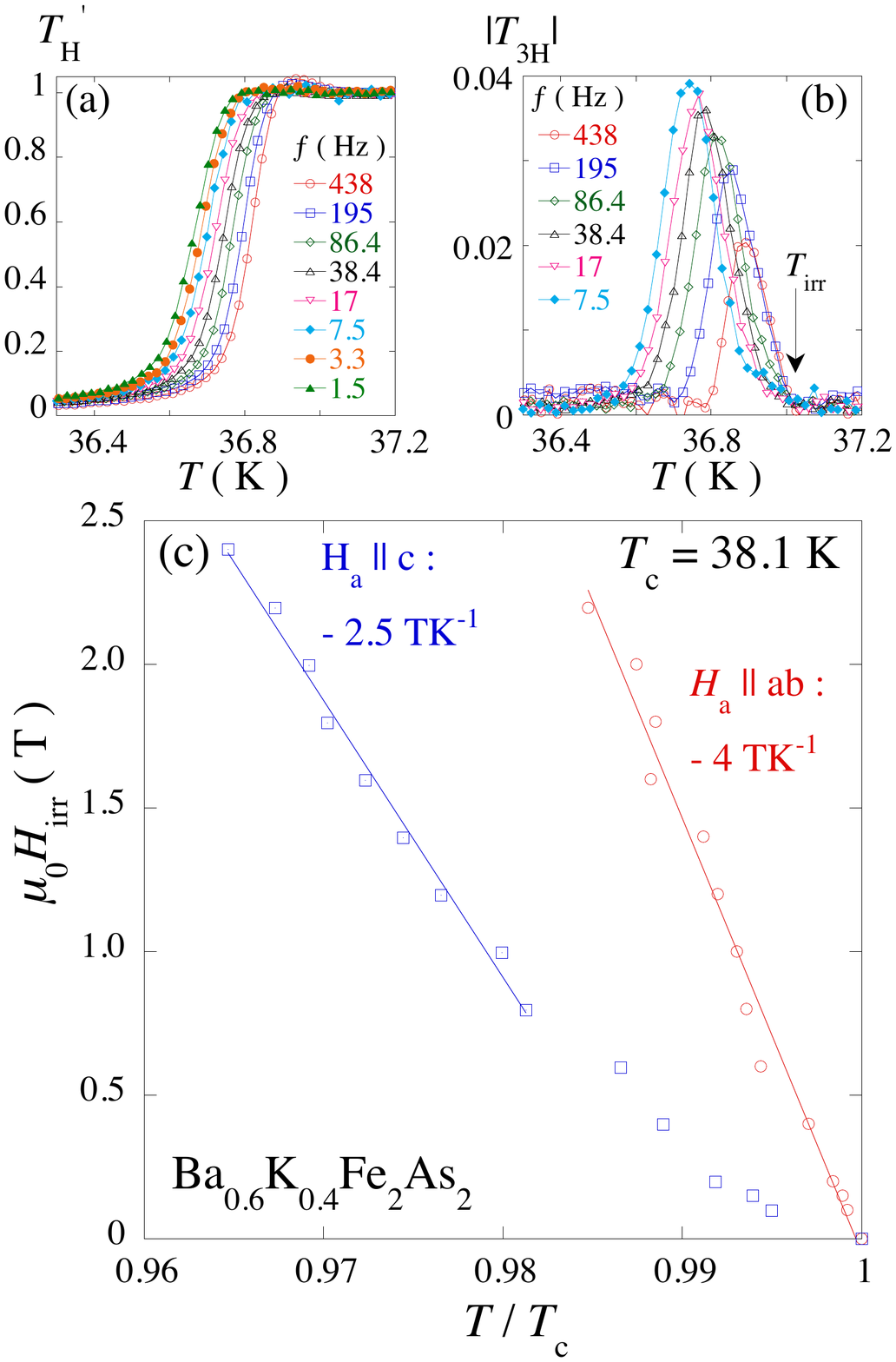}%
\vspace{-20mm}
\caption{(Color online)  (a,b) In-phase fundamental ($T^{\prime}_{H}$) and amplitude of the third harmonic ac transmittivity ($|T_{H3}|$) as function of the temperature, at an applied field $\mu_{0}H_{a} = 1.5$~T, for different indicated frequencies of the ac field.  (c) Irreversibility field $H_{irr}$, as determined from the vanishing of $|T_{3H}|$, indicated by the arrow in (b), for $H_{a} \parallel c$ and $H_{a} \parallel ab$ respectively.}
\label{IRL}%
\end{figure}

\section{Experimental details}

The Ba$_{0.6}$K$_{0.4}$Fe$_{2}$As$_{2}$ single crystals with $T_{c} \sim 38.1$ K (Fig.~\ref{IRL}) were grown by the self-flux method, using FeAs as the
self-flux. Details of the growth can be found in Refs.~\onlinecite{Luo2008,Shan2011}. For the experiment, crystals were cut to regular rectangles of dimensions $350 \times 160 \times 70 $ $\mu$m$^{3}$ using a wire saw. The spatial distribution of the local induction perpendicular to the surface of the crystals in the critical state was measured using an array of microscopic Hall sensors, fashioned in a pseudomorphic GaAlAs/GaAs heterostructure using ion implantation. The 10 Hall sensors of the array, spaced by 20 $\mu$m,  had an active area of $3 \times 3$ $\mu$m$^{2}$, while an 11$^{\rm th}$ sensor was used for the measurement of the applied field. For measurements with field parallel to $ab$, an array of sensors spaced by 10~$\mu$m was used. The Hall sensor array was placed on the center of the crystal surface, perpendicular to the long crystal edge, and spanning the crystal boundary. In this manner, hysteretic loops of the spatially resolved local induction were measured as function of the applied magnetic field $H_{a}$, and as function of temperature. The sweep-rate of the applied magnetic field was 40 G/s ($\mu_{0}\dot{H}_{a} = 4$ mT/s). In all experiments, the applied magnetic field was not only much smaller than the upper critical field $H_{c2}$, but also smaller than the field at which the second magnetization peak occurs.\cite{secondpeak}

\begin{figure}[t]%
\vspace{0mm}
\includegraphics[width=1.28\linewidth]{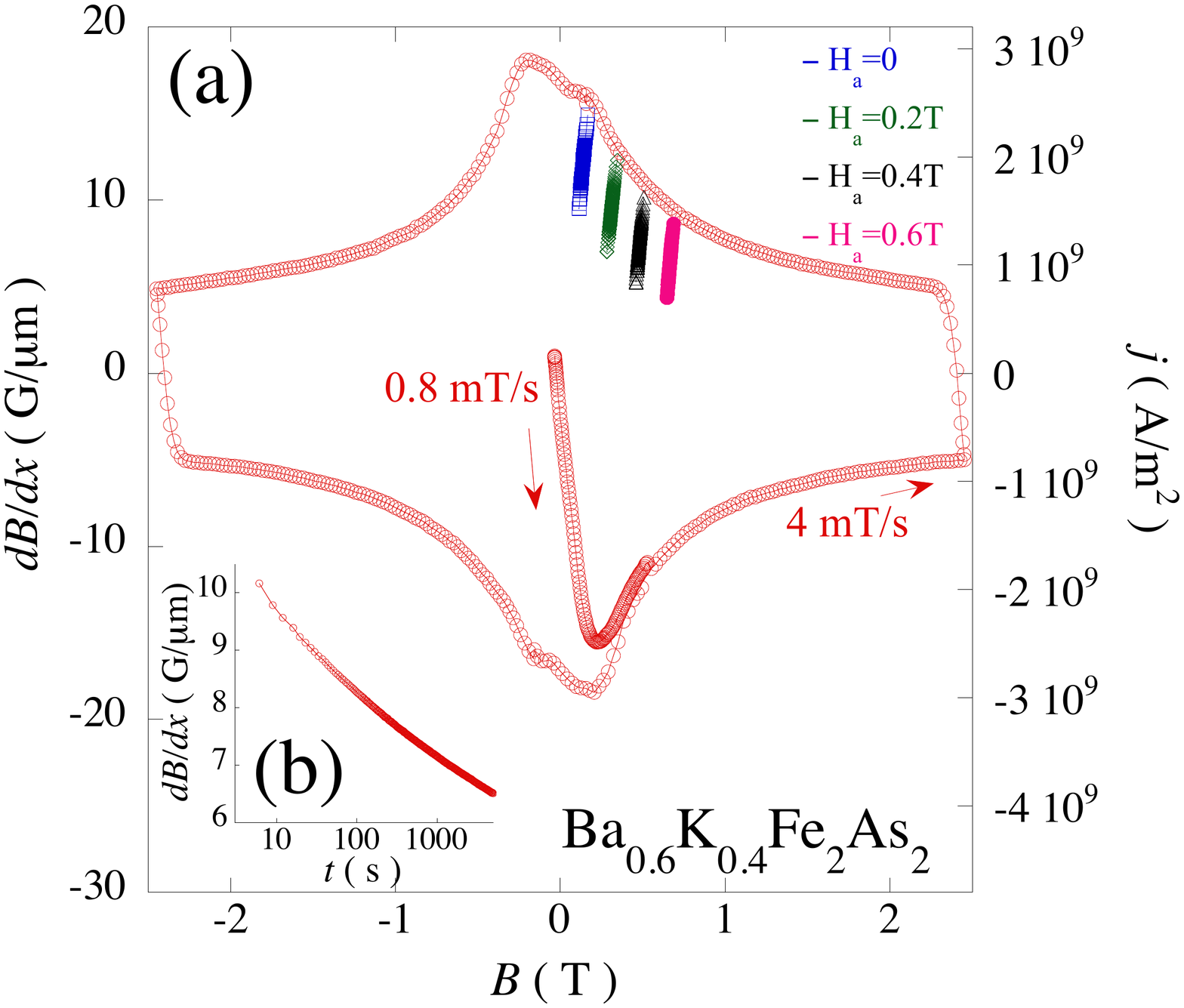}%
\vspace{-0mm}
\caption{(Color online)  (a) (\color{red}$\circ$\color{black}) Hysteresis loop of the gradient   $dB/dx \propto j_{ab}^{c}$ versus the local induction $B$, measured perpendicular to the long edge of the  Ba$_{0.6}$K$_{0.4}$Fe$_{2}$As$_{2}$ single crystal, for $H_{a} \parallel c$ at $T = 10$ K. The sweep-rate of the applied magnetic field was 0.8 mT/s for the virgin magnetization loop (up to $\mu_{0}H_{a} = 0.5$~T), while further measurements were carried out with a sweep-rate of  4 mT/s.  The four measurement sequences ( \color{blue}$\Box$\color{black},  \color{green}$\Diamond$\color{black}, $\triangle$,  \color{magenta}$\bullet$\color{black} ) illustrate the decay of the flux gradient at different applied fields, over a period spanning 5 to 5000 s after field-cooling, and reduction of the external field $H_{a}$ to the measurement field $H_{a} - \Delta H_{a}$ (with $\Delta H_{a} = 0.2$~T). (b). Time dependence of the local induction gradient for an applied field of 0.6 T and $T = 8$ K.}
\label{fig:creep}%
\end{figure}

\begin{figure}[t]%
\vspace{0mm}
\includegraphics[width=1.45\linewidth]{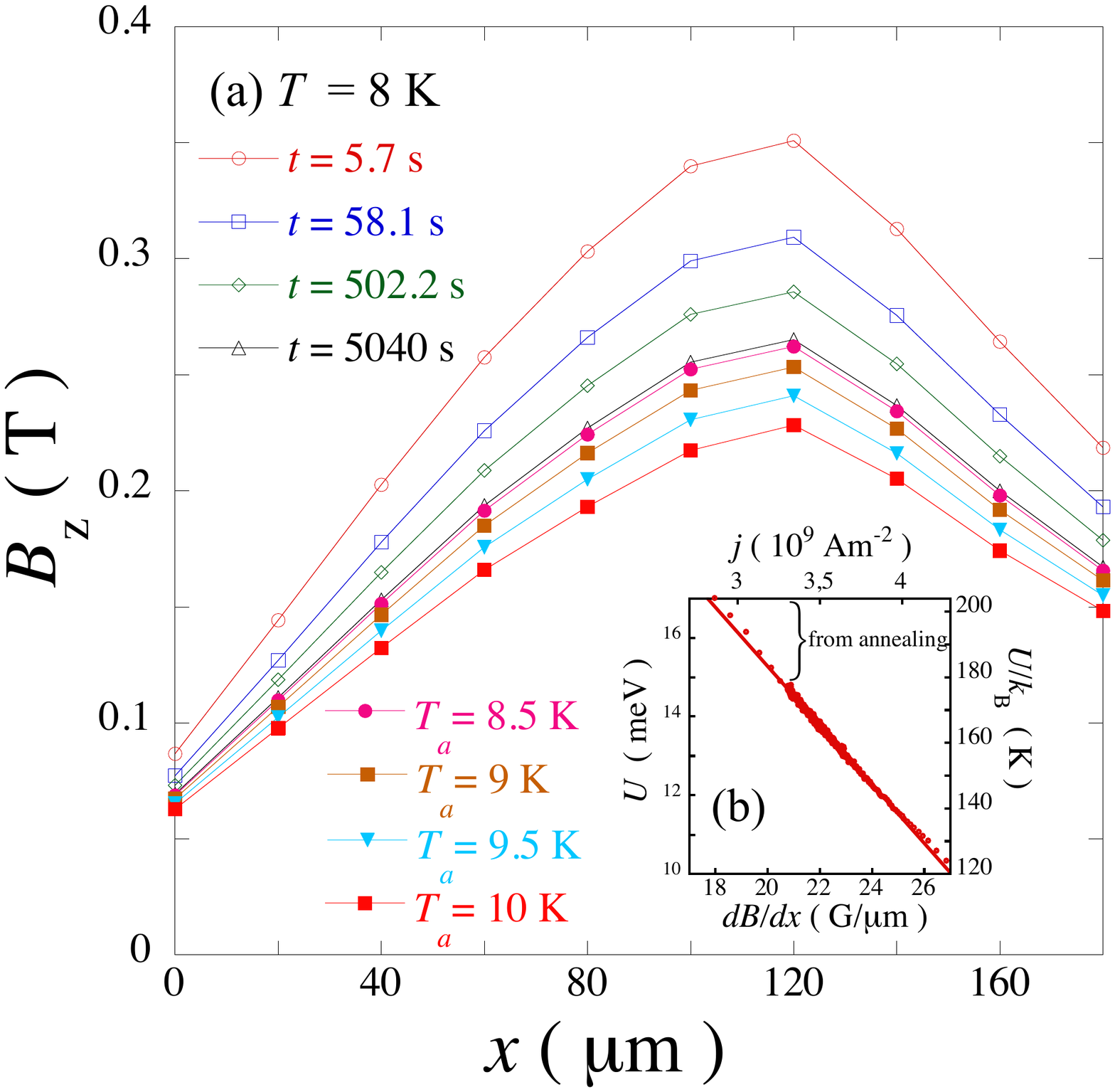}%
\vspace{-0mm}
\caption{(Color online)  (a) Relaxation of the critical state flux profile in the Ba$_{0.6}$K$_{0.4}$Fe$_{2}$As$_{2}$--crystal, for $H_{a} \parallel c$, at $T = 8$ K.  Open symbols show the flux density profile across the  Ba$_{0.6}$K$_{0.4}$Fe$_{2}$As$_{2}$ single crystal for four successive times, while closed symbols illustrate the flux density profiles obtained after the annealing of the critical state to the indicated temperatures. (b). Flux-creep activation barrier as function of  the screening current density, as extracted using Eq.~(\protect\ref{eq:Maley}).  }
\label{fig:profile}%
\end{figure}

Under all circumstances, the profiles of the flux density $B$  were well described by the critical state model,\cite{Bean62,Zeldov94} see Fig.~\ref{fig:profile}, allowing for the straightforward extraction of the local screening current density $j \approx (2/\mu_{0})dB/dx$. The factor $2$ takes the finite dimensions of the crystal into account; namely, the field gradient on the end surface of a semi-infinite bar in the critical state is half the field gradient in the interior.\cite{Brandt98}  Measurements were done with the Hall array parallel to the $ab$--plane and  $H_{a} \parallel c$, a configuration that measures the usual screening current density in the $ab$-plane, $j = j_{ab}^{c}$, and with the array $\parallel c$ for $H_{a} \parallel ab$, which yields the $ab$-plane screening current density $j_{ab}^{ab}$ corresponding to vortices moving parallel to the $c$-axis  (perpendicular to the FeAs layers).

Measurements of the ac screening were performed with the same set-up. A sinusoidally time-varying field of magnitude 1 Oe and frequency $f$ is applied colinearly with the dc field. The ac component $B_{ac}(f,T)$ of the local induction is then measured using the Hall probe array. Results, presented as the in-phase fundamental ac transmittivity $T_{H}^{\prime} = [B_{ac}(f,T) - B_{ac}(f,T\ll T_{c})]/[B_{ac}(f,T\gg T_{c}) - B_{ac}(f,T\ll T_{c})]$  and the third harmonic $|T_{3H}| = B_{ac}(3f,T)/[B_{ac}(f,T\gg T_{c}) - B_{ac}(f,T\ll T_{c})]$,\cite{Gilchrist93} are shown in Figs.~\ref{IRL}a and \ref{IRL}b respectively. The non-zero value of the latter signals the existence of a non-zero critical current density $j_{c}$; the vanishing of $B_{ac}(3f)$ at high-temperature is used to trace the dc irreversibility field $H_{irr}(T)$, for dc field aligned along the $c$--axis and the $ab$--plane respectively (Fig.~\ref{IRL}c).

As for the magnetic relaxation experiments, these were carried out on the decreasing field branch magnetization branch only (corresponding to flux exit relaxations), in order to prevent possible influence of surface barrier relaxation. The external magnetic field was applied at a temperature $T > T_{c}$, at which the Hall probe array was calibrated (with respect to the applied field). The crystal  was subsequently field-cooled  to the measurement temperature $T_{e} < T_{c}$, and the field reduced by an amount $\Delta H_{a}$ to the measurement field. Care was taken that $\Delta H_{a}$ exceeded the the field of full flux penetration, so that a full critical state is established. After waiting 5 s to allow for the settling of the magnet,  the flux density values at the different Hall sensor positions were measured as function of time, for a period of 5000 s. Furthermore, flux creep annealing experiments\cite{Maley90} were performed by heating the sample to $T_{a} > T_{e}$,  returning to the experimental temperature $T_{e}$, and re-measuring $S$ over a period of 600 s. This procedure is equivalent to performing measurements at the effective time $t = \tau \exp[U(j(T_{a}))/k_{B}T_{e}]$, of the order $10^{6}$~s.

  \begin{figure}[b]%
\vspace{-3mm}
\includegraphics[width=1.35\linewidth]{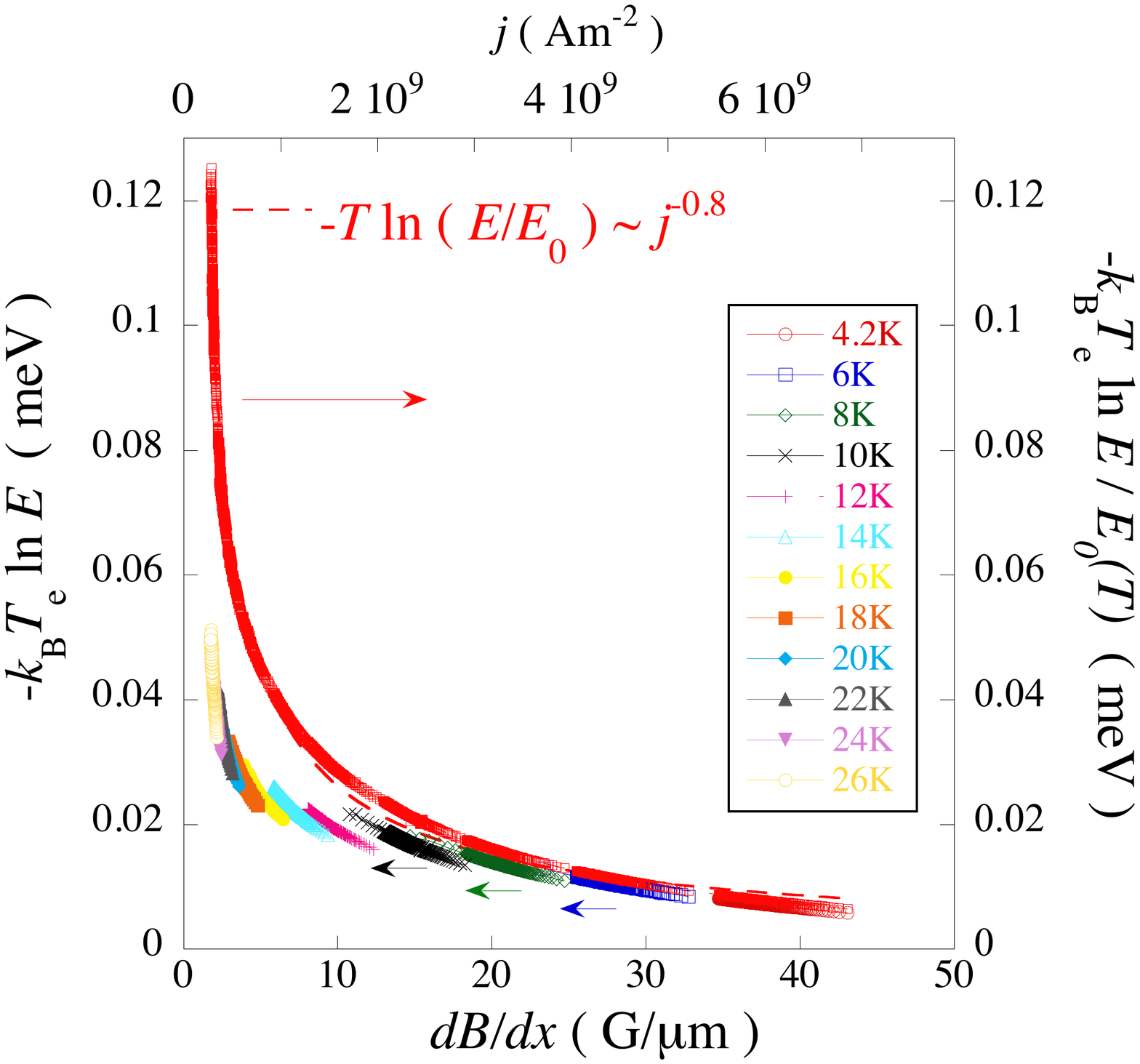}%
\vspace{-2mm}
\caption{(Color online)  Treatment of creep data for $\mu_{0}H_{a} = 0.2$ T $\parallel c$, using the method of Ref.~\protect\onlinecite{Abulafia95}. The data at the bottom of the frame show the $j$-dependence of the product of $T_{e}$ and the logarithm of the  electric field, such as extracted using Eq.~(\protect\ref{eq:E}), for different experimental temperatures $T_{e}$ (lefthand axis). The master curve at the top (righthand axis) was compiled from these data using a temperature-- and current density--independent $E_{0}$; however, as in Ref.~\protect\onlinecite{Abulafia95}, $E_{0}$ was assumed to be proportional to $B$; this yields a $\mu$--value of 0.8. }
\label{fig-Abulafia}%
\end{figure}

\section{Results}

 Fig.~\ref{fig:creep}(a) shows hysteresis loops of the flux density gradient $dB/dx$ as function of the local induction $B$ for $H_{a} \parallel c$, at $T = 10$ K. The width of the loop is proportional to the screening current density $j_{ab}^{c}$ in the $ab$-plane,  for field $\parallel c$. It has the characteristic shape found in all charge--doped iron-based superconductors:\cite{Kees} a ``central peak'' of $j$ around $B = 0$, followed by a drop $j \propto B^{-1/2}$ characteristic of strong vortex pinning by nm-scale point defects\cite{Kees1,Ovchinnikov91,vdBeek2002} or heterogeneities. As $B$ increases, the strong pinning-contribution to the critical current becomes irrelevant, and $j(B)$ saturates to a field--independent value determined by weak collective pinning of individual vortices,\cite{Blatter94} presumably by the K dopant atoms in the material.\cite{Kees} The magnitude of $ j_{ab}^{c}$ is very similar to that found in other measurements on the same material.\cite{Yang2008,Wang2010,Kees}

 At all magnetic fields, the sustainable current relaxes as function of time, with $S \sim -0.06$ [Fig.\ref{fig:creep}(b)]. Creep is logarithmic in time, with curvature indicative of a nonlinear $U(j)$ relation. However, contrary to the cuprate superconductors, and as illustrated by the near--linear evolution of the experimentally determined flux-creep activation barrier in the inset to Fig.~\ref{fig:profile}, the accessed dynamical range of $S(j(t))$ is too small in the iron-based superconductors to reliably extract a value of $\mu$ directly.  The effect of flux creep-annealing is also illustrated in Fig.~\ref{fig:profile}, which shows the relaxation of the trapped flux-profile at $T = 8$ K together with profiles obtained after annealing at various temperatures. 
 
 The time--dependent current density is analyzed using the method of Maley {\em et al.}.\cite{Maley90} This  yields the experimental current density-dependent activation barrier as 
\begin{equation}
 U_{e}(j) = - k_{B}T_{e} \ln\left( \left|\frac{dj}{dt}\right| \right) + c T_{e} , 
\label{eq:Maley}
\end{equation}
with $c$ a temperature-independent constant, to be chosen so that segments corresponding to the $U_{e}(j)$ relation measured at different temperatures line up to form the $U(j)$ relation representative of the vortex creep mechanism governing the magnetic relaxation. For the collective creep mechanism of Ref.~\onlinecite{Feigelman89}, 
\begin{equation}
j = \frac{j_{c}}{\{(k_{B}T/U_{c}) \ln\left[ \left(t_{0}+t \right)/\tau\right]\}^{1/\mu}},
\label{eq:j(T)}
\end{equation}
so that 
the method gives\cite{vdBeek92II}  
\begin{equation}
- k_{B}T_{e} \ln\left(\left |\frac{dj}{dt}\right| \right) = U(j)  - k_{B}T_{e}\ln\left[ \frac{k_{B}T_{e}}{U(j)} \frac{j}{\mu\tau} \right];
\end{equation}
hence, $c \equiv k_{B}\ln\left[ k_{B}T_{e}j /U(j) \mu\tau \right]$ actually depends logarithmically on temperature. For a putative logarithmic dependence $U =U_{c}\ln(j_{c}/j)$,\cite{Vinokur91,GurevichBrandt94} yielding $ j = j_{c} \left[ \left( t_{0}+t \right) / \tau \right] ^{-k_{B}T/U_{c}}$, one has\cite{vdBeek92II}  
\begin{equation}
- k_{B}T_{e} \ln\left( \left|\frac{dj}{dt} \right|\right) = U(j)  - k_{B}T_{e}\ln\left( \frac{k_{B}T_{e}}{U_{c}} \frac{j_{c}}{\tau} \right).
\end{equation}
The following shortcomings of the method of Ref.~ \onlinecite{Maley90} are identified. While, at low temperature, only a single $c$-value will satisfy the requirement of lining up the measured segments, at higher $T$ the temperature dependence of $c$ becomes important. Second, the prefactor $U_{c}(T) \equiv U_{c}(0)f(T)$, as introduced in Eqs.~(\ref{eq:cc}) and (\ref{eq:interpolation}), itself introduces a more important temperature dependence. This  can be recovered by dividing the result of Eq.~(\ref{eq:Maley}) at each $T_{e}$ by phenomenological factors $f(T_{e})$, with, again,  a degree of arbitrariness. Correcting for the temperature dependences of $c$ and $U_{c}$ in different ways results in different final results for the compiled $U(j)$--curve. Finally, the method supposes that the same $U(j)$ mechanism governs flux creep at all the  temperatures used to reconstitute the experimental $U(j)$--curve. In the present set of experiments, we find that applying different procedures to cope with these different temperature dependences yield an error bar on $\mu$ that amounts to 50 to 100~\% of its value.

\begin{figure}[t]%
\vspace{-12mm}
\includegraphics[width=1.5\linewidth]{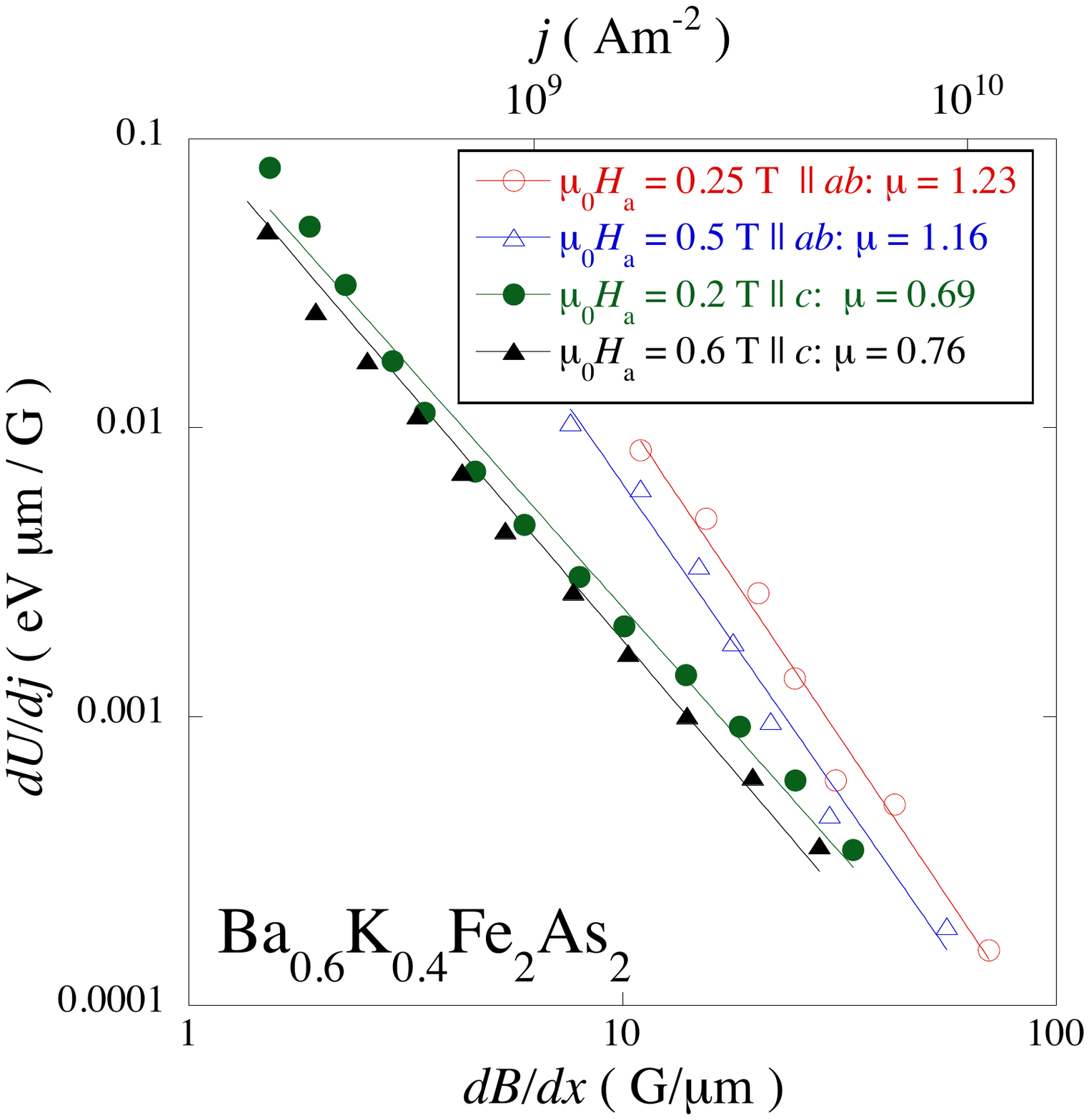}%
\vspace{-2mm}
\caption{(Color online) Compiled double-logarithmic plot of $\langle d[ -  k_{B}T_{e} d\ln(|dj/dt|)]/dj \rangle$ vs.  $\langle j \rangle$, as suggested from Eq.~(\protect\ref{eq:derivative}), for all temperatures, and the two orientations ($\parallel c$ and $\parallel ab$) of the applied magnetic field.}
\label{fig3}%
\end{figure}

Because it involves the integral (\ref{eq:E}) rather than the local value $dj/dt$, the method of Ref.~\onlinecite{Abulafia95} has the merit of yielding more accurate $U(j)$--values. In our experiments, we determine the electric field by integrating to the sample boundary; in accordance,  the relevant value of the current density is that which corresponds to the slope $dB/dx$ at the  boundary.  Fig.~\ref{fig-Abulafia} shows $-k_{B}T \ln E$ thus determined, for various temperatures, and $\mu_{0}H_{a} = 0.2$ T $\parallel c$. However, if one wishes to extract flux-creep activation energies, one is faced with the same arbitrariness concerning the factor $c$ in Ref.~\onlinecite{Maley90}, now contained by the factor $E_{0}$.  Compiling a full $U(j)$--curve using a temperature- and current-density independent $E_{0}$ yields the illustrated master curve, suggesting that $U(j) \propto j^{- \mu}$ with $\mu = 0.8$. However, if one takes $E_{0}$ to be proportional to $Bj$, as in Ref.~\onlinecite{Abulafia95}, one obtains a curve that fits $U(j) \propto j^{-  0.25}$.

To avoid ambiguities, we evaluate the averaged current density-derivative 
$\langle -  k_{B}T_{e} d \ln|E|/dj \rangle$ 
 at each $T_{e}$. This procedure has the advantage of eliminating the prefactor $c$ (or, equivalently, $E_{0}$) from the analysis. For Eq.~(\ref{eq:cc}), one has 
\begin{eqnarray}
\langle  \frac{ k_{B}T d\ln|E|}{dj} \rangle & = & - \langle \frac{(\mu + 1) k_{B}T_{e}}{j} - \frac{dU(j)}{dj}\rangle  \\
& \approx & \mu \frac{U_{c}}{j_{c}} \left( \frac{j_{c}}{\langle j\rangle } \right)^{1+\mu}, \hspace{0mm} (U_{c} \gg k_{B}T) \nonumber
\label{eq:derivative}
\end{eqnarray}
while for power-law creep\cite{Vinokur91,GurevichBrandt94} 
\begin{equation}
\langle  \frac{k_{B}T d \ln |E|}{dj} \rangle = \frac{U_{c}}{\langle j \rangle}.  \hspace{15mm} (U_{c} \gg k_{B}T)
\label{eq:log-derivative}
\end{equation}
A double-logarithmic plot (Fig.~\ref{fig3}) of the average  
 $\langle -  k_{B}T_{e} d\ln|E|/dj \rangle$ vs.  $\langle j \rangle$ for all different temperatures $T_{e}$ therefore directly reveals the creep mechanism, as well as the value of $\mu$, from the deviation of the linear slope from 1. Whether the condition $U_{c} \gg k_{B}T$ is verified can be easily checked from the Bean (straight-line)--like nature of the flux profiles in Fig.~\ref{fig:profile}.\cite{Gilchrist94}

\begin{figure}[t]%
\vspace{3mm}
\includegraphics[width=1.3\linewidth]{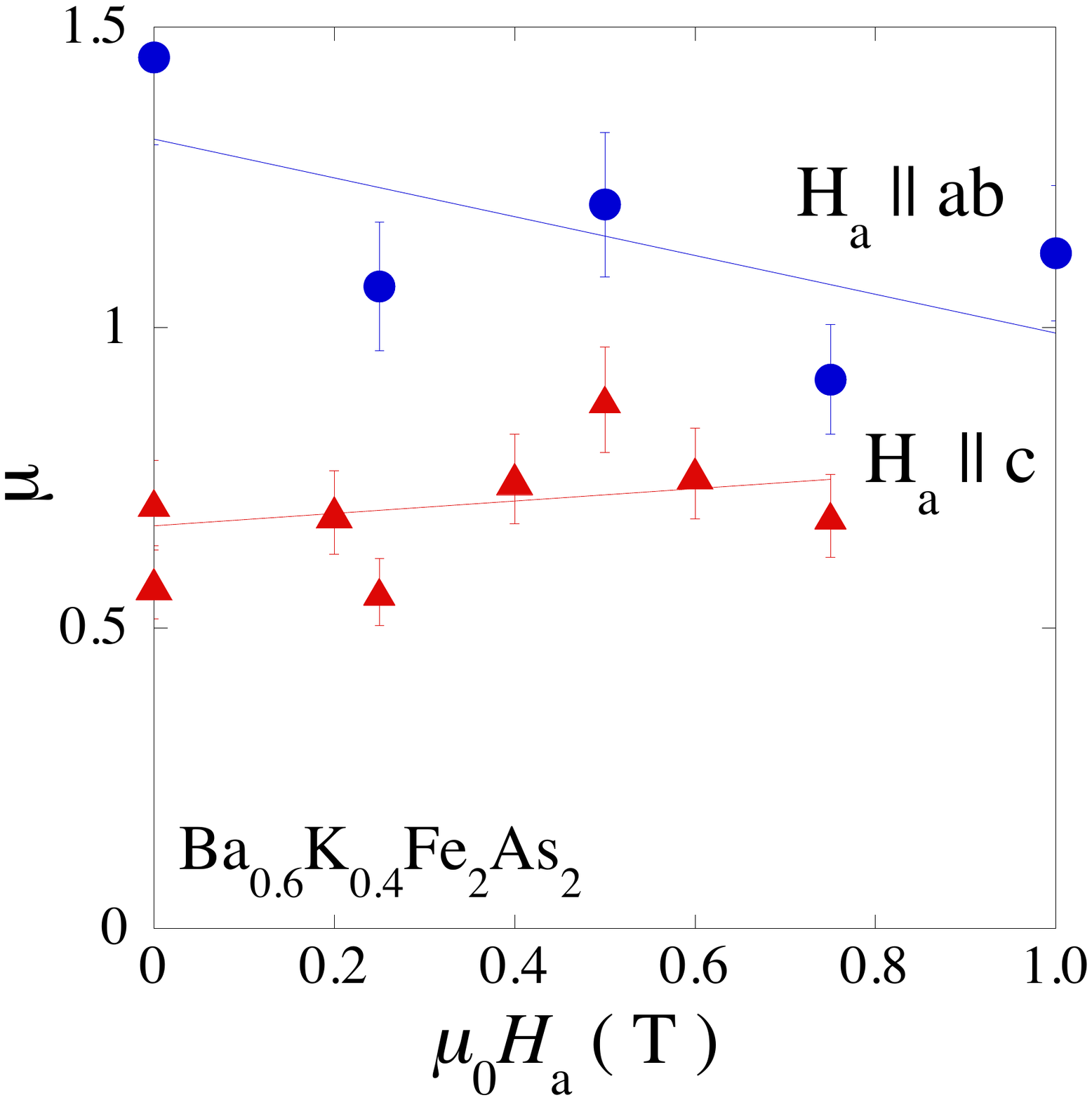}%
\vspace{-7mm}
\caption{(Color online) Experimental values of the creep exponent $\mu$ for the two orientations of the applied magnetic field.}
\label{fig-mu}%
\end{figure}

From the slopes in Fig.~\ref{fig3}, we find that Eqs.~(\ref{eq:cc}) and (\ref{eq:interpolation}) describe the data satisfactorily. The corresponding $\mu$--values are rendered in Fig.~\ref{fig-mu}. For $H_{a} \parallel c$, $\mu = \mu^{c} \approx 0.65$ at low fields, slowly increasing towards $ \mu^{c} \approx 0.8$ at higher fields, while for $H_{ab} \parallel ab$, $\mu^{ab} \approx 1.5$ at low fields, decreasing towards $\mu^{ab} \approx 1$ at higher fields.  The results for $H_{a} \parallel c$ are remarkably similar to those found in Ba(Fe$_{0.92}$Co$_{0.08}$)$_{2}$As$_{2}$ by Shen {\em et al.}.\cite{BingShen2010}

\begin{figure}[t]%
\includegraphics[width=1.45\linewidth]{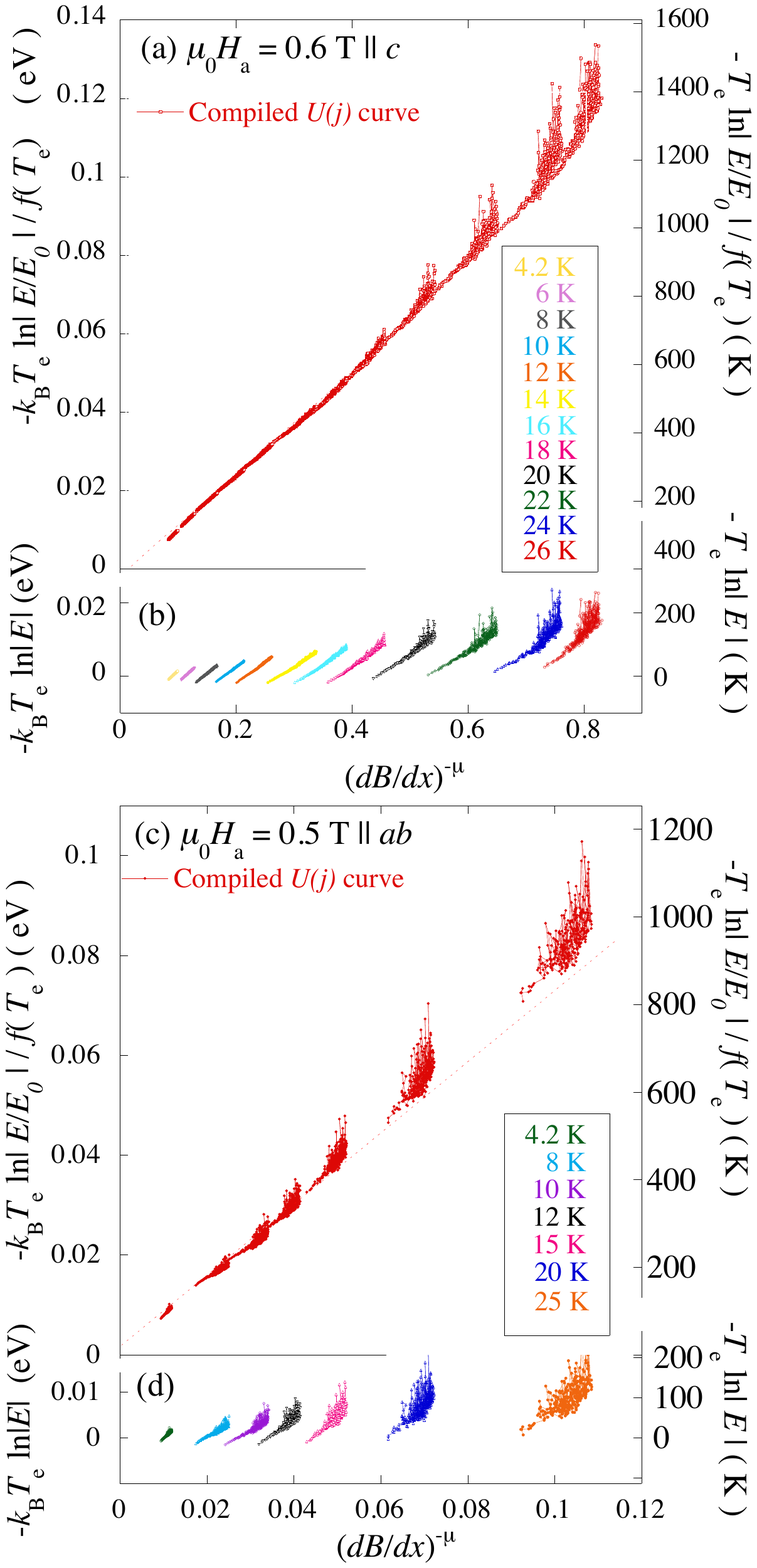}%
\vspace{-20mm}
\caption{(Color online) Current-density dependence of the barrier $U(j)$ for thermally activated vortex creep in the Ba$_{0.6}$K$_{0.4}$Fe$_{2}$As$_{2}$ crystal, for (a,b)  $\mu_{0}H_{a} = 0.6$ T  $\parallel c $, $\mu = 0.75$; and (c,d)  $\mu_{0}H_{a} = 0.5$ T  $\parallel ab $, $\mu = 1.13$. (b) and (d) show the (logarithm of) the raw electric field data, multiplied by the experimental temperature, $ -  k_{B}T_{e} d\ln|E| $, versus $(dB/dx)^{-\mu} \propto j^{-\mu}$. (a,c) show the full, compiled curves of the creep barrier, obtained by adding the relevant factor $k_{B}T_{e} \ln E_{0}$ for each temperature. }%
\label{fig:Maley}%
\end{figure}

\begin{figure}[t]%
\includegraphics[width=1.15\linewidth]{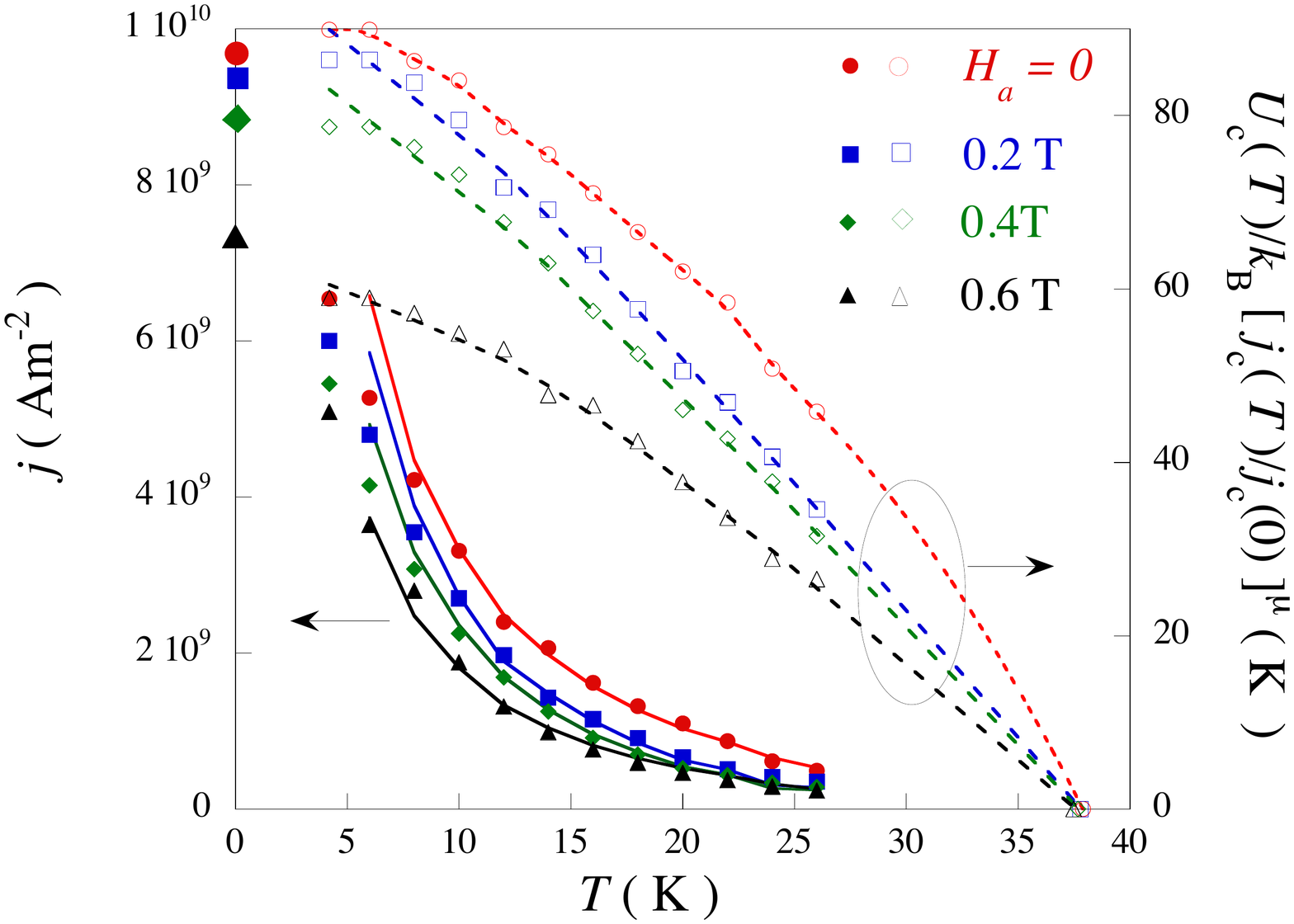}%
\caption{(Color online) Comparison of the temperature dependence of the activation barrier $U_{c}(T) [j_{c}(T)/j_{c}(0)]^{\mu}$ (open symbols) extracted from the creep experiments, with that of the measured screening current density $j(T)$ at the onset of relaxation (closed symbols). The $j(T)$ values (closed symbols) are tantamount to those one would measure in magnetic hysteresis experiments. Large closed symbols on the lefthand abscissa denote the  $j_{c}( T = 0)$--values, determined from the extrapolation of the activation barrier to zero in Fig.~\protect\ref{fig:barrier}.  Drawn lines denote Eq.~(\protect\ref{eq:j(T)}), evaluated using the experimental values of the product $U_{c}j_{c}^{\mu}$, and $\ln(t_{0}+t/\tau) = 21$. Dashed lines are guides to the eye.}%
\label{fig:U(T)}
\end{figure}

\begin{figure}[t]%
\includegraphics[width=1.3\linewidth]{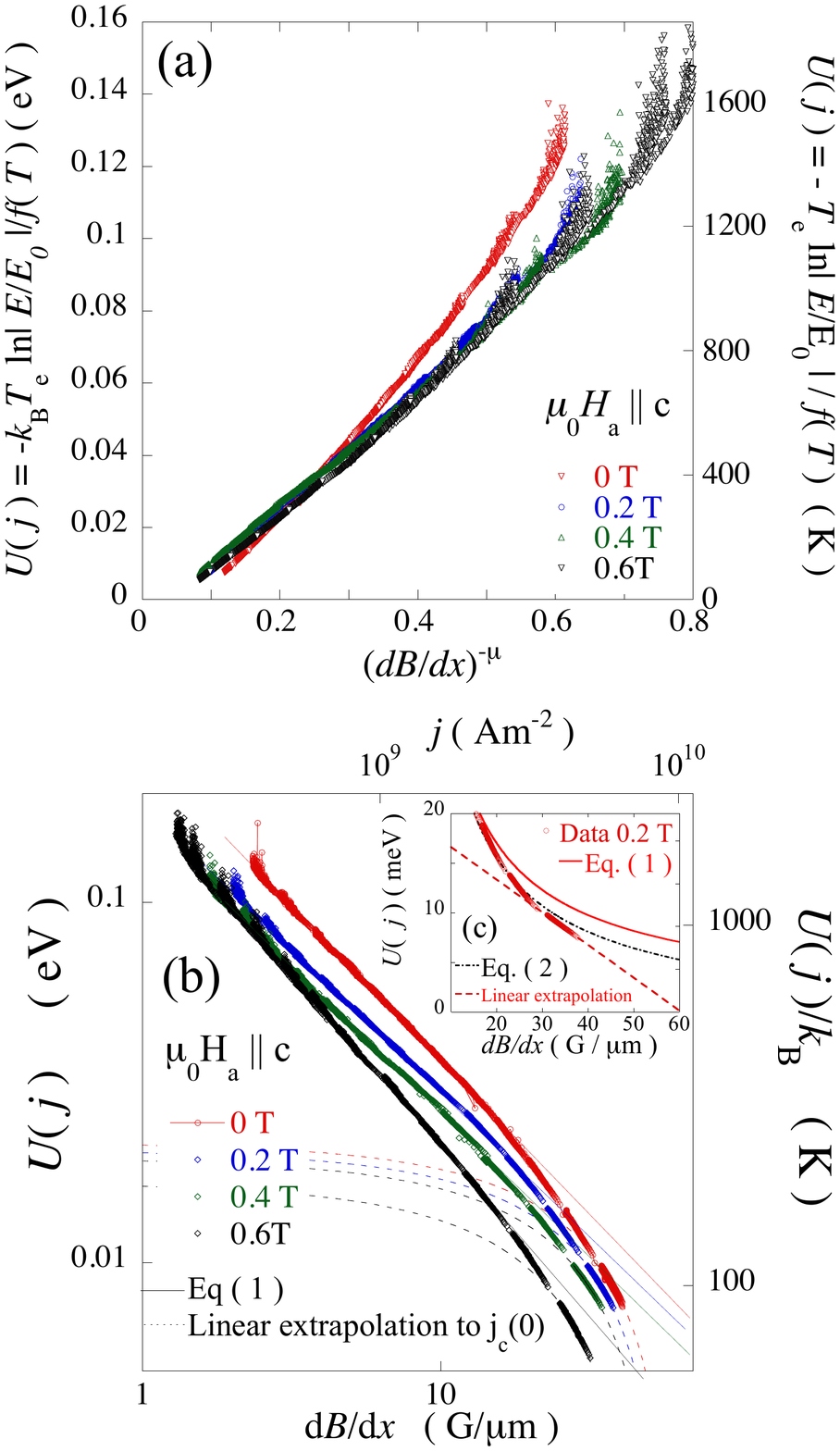}%
\caption{(Color online) (a) Plot of $ -  k_{B}T_{e} d\ln|E| $ vs. $(dB/dx)^{-\mu} \propto j^{-\mu}$, for different magnetic fields strengths $\parallel c$. (b)  Flux-creep activation barrier compiled from relaxation experiments at different temperatures, for the same fields as (a). The dotted lines show the linear extrapolation of the low-temperature barrier to  $j_{c}(0)$. The drawn lines show fits to the collective creep theory, Eq.~(\ref{eq:cc}). (c) Data for applied field $\mu_{0}H_{a} = 0.2$~T, showing the inadequacy of the interpolation formula (\protect\ref{eq:interpolation}) (dashed-dotted line). }%
\label{fig:barrier}%
\end{figure}

The obtained $\mu$--values are checked by plotting 
$ -  k_{B}T_{e} \ln|E|$ vs. $ j^{-\mu}$ for all temperatures and fields. Fig.~\ref{fig:Maley} shows that this yields straight lines for both field orientations, as required. Deviations from linearity are only apparent below 8 K, and above 24 K, which means that $\mu$ is $T$--independent for the greater part of the investigated temperature range.  That is, the same mechanism governs vortex creep for all $T < 24$ K. The slopes in Fig.~\ref{fig:Maley} yield $U_{c}j_{c}^{\mu}$, while the intercept with the abscissa corresponds to $j_{e}^{-\mu}$, where $j_{e} = j_{c}\{ U_{c}/ k_{B}T_{e} \ln\left[k_{B}T_{e} j /U(j) \mu \tau \right]\}^{1/\mu}$. With the uncertainty regarding $\mu$ removed, the curves obtained for different $T_{e}$ can now be compiled into a ``zero-temperature'' curve by adding factors $k_{B}T_{e}\ln E_{0}$, and dividing by  $f(T_{e})$. For the lowest two temperatures, we set $f(T_{e}) = 1$ to obtain an unequivocal  $E_{0} = 0.019$. For higher $T_{e}$, the value of  $f(T_{e})$ is adapted in order to line up the relevant data segments to obtain a continuous compiled curve, with continuous derivative.  The result is given in Fig.~\ref{fig:barrier}, which shows the extracted activation barrier $U(j)$ as function of $j^{-\mu}$ (top panel) and as function of $j$ (bottom panel), for various magnetic fields $\parallel c$. The extracted temperature dependence $U_{c}j_{c}^{\mu}$ for $H_{a} \parallel c$ is depicted in Fig.~\ref{fig:U(T)}.

\section{Discussion}

Inspection of the creep barrier in Fig.~\ref{fig:barrier} shows deviations from power-law behavior at low temperature < 8 K, and above 24 K. In all cases, the dependence $U(j)$ at the lowest temperature appears to be \em linear \rm in the current density $j$. This behavior is interpreted as being due to the proximity of the measured screening current density to the (pinning) critical current density. A linear extrapolation of the $U(j)$ curves to zero (dotted lines in Fig.~\ref{fig:barrier}) thus provides an estimate of the critical current density  $j_{c}(0)$ in the limit $T\rightarrow 0$. The obtained values range between $j_{c}(0;0.6 {\,\mathrm T}) = 6.7 \times 10^{9}$ Am$^{-2}$ to  $j_{c}(0;H_{a} = 0) = 9.7 \times 10^{9}$ Am$^{-2}$ for field $\parallel c$, and  $j_{c}(0;0.5 \,{\mathrm T})  \approx  2.3 \times 10^{10}$ Am$^{-2}$ for $H_{a} \parallel ab$ (Fig.~\ref{fig:U(T)}).   These values can be factored out from the slopes of the curves in Fig.~\ref{fig:barrier}(a,b) to yield  creep barrier values of $U_{c}(0)/k_{B} = $ 60 -- 85 K, decreasing with increasing values of field $\parallel c$, see Fig.~\ref{fig:U(T)}, and $U_{c}(0)/k_{B} = 24$ K for $H_{a} \parallel ab$. The low-temperature values of the measured screening current density $j$ closely approach $j_{c}(0)$. These values therefore yield a good order-of-magnitude estimate of the bulk pinning force.

The Inset, Fig.~\ref{fig:barrier}(c), compares the high-- and low current limiting behavior of the current density--dependent creep barrier for $\mu_{0}H_{a} =  0.2$ T $\parallel c$ to the often--used interpolation formula, Eq.~(\ref{eq:interpolation}).  It is clear that the current density range over which the experimental barrier crosses over from the high--current, linear-in-$j$, to the low current--behavior given by Eq.~(\ref{eq:cc}) is much narrower than what can be described using Eq.~(\ref{eq:interpolation}). In fact, the dashed--dotted curve shows that applying the interpolation formula (\ref{eq:interpolation}) yields a gross overestimate of the pinning critical current density, as well as a possible overestimate of $\mu$. It therefore seems imperative to use Eq.~(\ref{eq:cc}), which was derived on physical grounds,\cite{Feigelman89} rather than the phenomenological formula (\ref{eq:interpolation}).

The extracted parameter values can be used to cross-check the analysis. The drawn lines in Fig. ~\ref{fig:U(T)} render a numerical evaluation of  Eq.~\ref{eq:j(T)} using the experimentally extracted $U_{c}j_{c}^{\mu}$ products, and $\ln( t_{0} + t / \tau ) = 21$. In the intermediate temperature range, at which creep is described by the barrier (\ref{eq:cc}), the agreement with the temperature--dependent screening current density values $j(T)$ at the onset of the relaxation (such as these might be measured during a field--sweep measurement) is more than satisfactory. Given that all times the current density is, to good approximation, given by the equation $U(j) = k_{B}T \ln( t_{0} + t / \tau )$,\cite{vdBeek92II,Schnack92,GurevichBrandt94} one has $\ln( t_{0} + t / \tau ) = U(j)/k_{B}T$, where the numerator and denominator can be simply read from Figs.~\ref{fig-Abulafia} and \ref{fig:barrier}. At the onset of relaxation ($t \sim 5$ s), one finds values ranging from 20.5 (for 4.2 to 12~K)  to 52 (at $T = 26$~K). Thus, the value $\ln( t_{0} + t / \tau ) = 21$ is  reasonable in the intermediate temperature range. The expression $\tau = (\Lambda j_{c} / E_{0})(k_{B}T/U_{c}) \sim 10^{-8}$~s allows one to estimate  $\ln(t_{0} + t / \tau )$ independently; for $t \sim 5$~s one again has a value of 20. For measurements performed with a commercial superconducting quantum interference device--based magnetometer, $t \sim 100$~s, so that $\ln( t_{0} + t / \tau  )$ is slightly larger. 

Fig.~\ref{fig:U(T)} shows that the field dependence of the screening current density, as expressed by Eq.(\ref{eq:j(T)}), is contained by the parameter $U_{c}j_{c}^{\mu}$. Hence, the $j(B)$--dependence is not the consequence of a field-dependence of the creep process, which would be reflected by a strongly field--dependent $\mu$; rather, it reflects the intrinsic field dependence of the pinning force. The $B^{-1/2}$--dependence of the screening current density in iron-based superconductors was recently interpreted in terms of strong pinning\cite{Ovchinnikov91,vdBeek2002} by nm-sized heterogeneities.\cite{Kees,Kees1,Sultan}
In this respect, the creep exponent $\mu \sim $ 0.6 --0.8 at low fields parallel to the $c$-axis, comparable to $\mu = 0.5$ found for single vortex creep in the Bragg-glass phase in single crystals of the cuprate high temperature superconductor Bi$_{2}$Sr$_{2}$CaCu$_{2}$O$_{8+\delta}$,\cite{Fuchs98III,vdBeek2000}  comes as somewhat of a surprise. Namely, vortex pinning at low fields in Bi$_{2}$Sr$_{2}$CaCu$_{2}$O$_{8+\delta}$ is thought to be not in the strong pinning limit, but in the opposite, weak pinning limit.\cite{Giamarchi95} In the field-values under scrutiny, the field--dependence observed in Fig.~\ref{fig:creep} suggests that one is dealing with strong pinning (see also Ref.~\onlinecite{SalemSugui}). A possible explanation for the similarity of the creep exponents in the two cases is that the velocity of the flux lines at low and intermediate currents is limited by the progression of roughened vortex segments spanning the region \em between \rm strong pins through the background of weak pins. This would seem natural given that weak pinning in the iron--pnictide superconductors is thought to be due to the \em local \rm fluctuations of the dopant atom density, while  strong pinning would originate from inhomogeneity of the dopant atom density on a much larger length scale. Thus, vortex segments would progress more or less continuously through areas of homogeneous doping, before becoming stuck on a ``strong pin'' , that is, a  region in which the dopant atom density averaged over several dozen nm significantly deviates from the overall mean. The critical current density, and the screening current density at low temperature, would then be determined by the strong pins, while creep at intermediate and high temperatures would be determined by the weak pinning background. An alternative hypothesis is that the creep exponent would be determined by the shape of the energy distribution of the vortices pinned by large--scale heterogeneity, much in the same way as this was proposed for  creep through columnar defects in the so-called variable--range hopping regime.\cite{Tauber95}


Finally, we remark that the consistently larger $\mu$--values found for $H_{a} \parallel ab$ are not unexpected, because, in the investigated orientation, $j \parallel ab$ but $\perp H_{a}$. Vortex lines are therefore oriented in the $ab$--plane, but are forced to move parallel to the crystalline $c$-axis. This is the hard direction for vortex motion, requiring the nucleation of vortex loops in the $c$-direction, a process limited by the value of the out-of-plane vortex lattice tilt modulus $c_{44}^{\perp} \sim \varepsilon_{\lambda}^{-3} \tilde{c}_{44}$. This exceeds the non-local tilt modulus $\tilde{c}_{44}$ for vortices $\parallel c$ by the inverse cube of the anisotropy factor $\varepsilon_{\lambda} \sim  0.4$.

\section{Summary and Conclusions}

Iron based-superconductors show a surprisingly large value of the relaxation rate $S$ of the irreversible magnetization. Since the relaxation rate is too small to directly extract  the current-density dependence of the flux-creep activation barrier from the time dependence of $1/S$, an  alternative, straightforward  analysis  method is proposed, that directly yields the relevant creep mechanism,  and the temperature dependence of the pinning parameters in the absence of creep. Applying this to single crystalline Ba$_{0.6}$K$_{0.4}$Fe$_{2}$As$_{2}$, we find evidence for nucleation-type (collective) flux creep, with a weakly field--dependent creep exponent, $\mu = $ 0.6--0.8 for magnetic fields oriented along the $c$-axis, and a slightly larger $\mu = 1.2 $ -- 1.5 for field along $ab$. Several hypotheses leading to such $\mu$--values, among which, the combined action of strong and weak pinning centers, and a non-trivial pinning energy distribution function are proposed. At low temperature, the screening current approaches the pinning critical current; therefore, meaningful information on flux pinning in the iron-pnictide superconductors can be directly extracted from  low-temperature ($T \lesssim 5$~K) magnetic hysteresis experiments.  The field dependence of the screening current density is found to arise from the underlying mechanism of pinning, and not from varying creep rates due to the flux creep process. Finally, it is found that the crossover between low--current and high--current behavior of the flux creep activation barrier is  poorly described by the so-called ``interpolation formula''.

\begin{acknowledgments}
We thank V. Mosser for providing the Hall sensor arrays. This work was supported by the French National Research agency, under grant ANR-07-Blan-0368 ``Micromag". The work at Ames Laboratory was supported by the U.S. Department of Energy, Office of Basic Energy Sciences, Division of Materials Sciences and Engineering under contract No. DE-AC02-07CH11358. Work at SNU was supported by National Creative Research Initiative (2010-0018300). The work of R. Prozorov in Palaiseau was funded by the St. Gobain chair of the Ecole Polytechnique. C.J. van der Beek and M. Konczykowski acknowledge the hospitality of Ames Lab and Iowa State University during the preparation of the manuscript. 
\end{acknowledgments}

\end{document}